\documentstyle[12pt]{article}
\begin{document}

\noindent
{\Large Lattice-Gas Models of Adsorption
in \vspace{0.4in} the Double
Layer}\footnote{Submitted to {\it Electrochimica Acta}.}

\noindent
{\Large Per Arne Rikvold$\;^{\rm a,b,}$
\footnote{Electronic address: rikvold@scri.fsu.edu.}$^,$
\footnote{Corresponding author.},
 Jun Zhang$\;^{\rm a}$,
 \vspace{-0.1in} Yung-Eun Sung$\;^{\rm c}$
\begin{center} and Andrzej \vspace{0.25in}
Wieckowski$\;^{\rm c}$ \end{center}}

\noindent $^{\rm a}$ Supercomputer
Computations Research Institute,
Florida State University, Tallahassee,
Florida \vspace{0.1in} 32306-4052, USA \\
$^{\rm b}$ Center for Materials Research and Technology,
and Department of Physics,
Florida State University, Tallahassee,
Florida \vspace{0.1in} 32306-3016, USA \\
$^{\rm c}$ Department of Chemistry and Frederick Seitz Materials
Research Laboratory,
University of Illinois, Urbana, Illinois 61801, USA

\begin{abstract}
The theory of statistical-mechanical lattice-gas modeling
of adsorption is reviewed and shown to be applicable to
a range of electrochemical problems dominated by effective, lateral
adsorbate--adsorbate interactions. A general strategy for applying
the method to specific systems is outlined, which includes
microscopic model formulation, calculation of zero-temperature phase
diagrams, numerical calculation of thermodynamical
and structural quantities at nonzero temperatures,
and estimation of effective, lateral
interaction energies that cannot be obtained by first-principles
methods. Phenomena that are discussed include poisoning and
enhanced adsorption, and illustrative applications to
specific systems
are reviewed. Particular problems considered are:
the poisoning by sulfur
of hydrogen adsorption on platinum (111),
the electrochemical adsorption
of naphthalene on polycrystalline copper and of urea
on single-crystal
platinum (100), and the underpotential deposition of
copper on single-crystal \vspace{0.1in} gold (111).

\end{abstract}

\section{Introduction}
\label{secIN}

Statistical-mechanical lattice-gas modeling provides a
paradigm for analyzing
site-specific single- and multicomponent chemisorption
at electrode--electrolyte interfaces. The
method is particularly useful to describe spatial ordering and
fluctuations in the contact-adsorbed layer,
which are strongly influenced by effective,
lateral adsorbate--adsorbate interactions.
The history of successful lattice-gas studies of phase
transitions at
solid--vacuum and solid--gas interfaces \cite{ZANG88}
makes the early
applications of the method to double-layer studies
\cite{RIKV88A,RIKV88B,COLL89,BLUM90,HUCK90,BLUM91,%
HUCK91,HUCK91B,ARMA91A,RIKV91A,RIKV91B,RIKV92}
excellent examples of the transfer
of a methodology from one research area to another.

Here we present a condensed review of the basics of
lattice-gas modeling
of specific adsorption in the double-layer region,
including a short discussion of poisoning and enhancement
effects and illustrated by
results from recent studies of specific systems.
The outline of the
remainder of the paper is as follows.
In Sec.~\ref{secMOD} we briefly review the
lattice-gas formulation
and some of the methods that can be used to obtain
specific numerical
results for such experimentally measurable
quantities as adsorption
isotherms, voltammetric currents and charge
densities, and
images obtained by low-energy electron diffraction
(LEED) and
atomic-resolution microscopies, such as scanning
tunneling microscopy (STM)
and atomic force microscopy (AFM).
In particular we concentrate on
non-perturbative numerical methods, such as
Monte Carlo (MC) simulations
\cite{COLL89,RIKV93A,RIKV93B,GAMB93B,HIGH93,RIKV95,%
JZHA95A,JZHA95B,BIND86,BIND92,BIND92B}
and transfer-matrix (TM) calculations
\cite{RIKV88A,RIKV88B,COLL89,RIKV91A,DOMB60,HUAN63,STAN71,NIGH90},
which are often combined with finite-size scaling methods
\cite{BIND92,NIGH90,PRIV90}.
In Sec.~\ref{secPE} we briefly
consider, within the lattice-gas picture,
such nonlinear effects in
multicomponent adsorption as poisoning
\cite{RIKV88A,RIKV88B,COLL89,ARMA91A,RIKV91B,RIKV92,HOMM89}
and enhanced adsorption
\cite{RIKV88B,ARMA91A,RIKV91A,RIKV91B,RIKV92,HOMM89,KRUK95},
both with semiquantitative applications to specific systems.
Reference \cite{RIKV91B} contains
more extensive discussions and comparisons of these
phenomena, which are just as relevant at
solid--vacuum and solid--gas
interfaces as they are in electrochemistry,
and which also can be
extended to multilayer adsorption \cite{KRUK95}.
In Sec.~\ref{secEX} we provide further quantitative
illustrations in the form
of applications to two specific cases of adsorption on
single-crystal electrodes: the electrosorption of
urea on Pt(100) from an acid electrolyte
\cite{RIKV92,RIKV93A,RIKV93B,GAMB93B,HIGH93,RIKV95,GAMB94}
and the underpotential deposition (UPD)
of copper on Au(111) from a sulfate-containing electrolyte
\cite{HUCK90,BLUM91,HUCK91,HUCK91B,JZHA95A,JZHA95B,%
BLUM93,BLUM94A,BLUM94B,LEGA95}.
A final summary and conclusions are given in Sec.~\ref{secDIS}.

\section{The Lattice-Gas Method}
\label{secMOD}

The lattice-gas models discussed here
are defined through a generalization of
the standard three-state
lattice-gas Hamiltonian (energy function) used, {\it e.g.}, in
Refs.~\cite{RIKV88A,RIKV88B,COLL89,RIKV91A,RIKV91B,RIKV92},
to give the energies of
particular adsorbate configurations:
\begin{eqnarray}
{\cal H}_{\rm LG}
&=& \sum_n \Bigg[ -\Phi_{\rm AA}^{(n)}
        \sum_{\langle ij \rangle}^{(n)} c_i^{\rm A} c_j^{\rm A}
       -\Phi_{\rm AB}^{(n)} \sum_{\langle ij \rangle}^{(n)}
        \left(c_i^{\rm A} c_j^{\rm B}
       + c_i^{\rm B} c_j^{\rm A} \right)
       -\Phi_{\rm BB}^{(n)}
       \sum_{\langle ij \rangle}^{(n)} c_i^{\rm B} c_j^{\rm B} \Bigg]
\nonumber\\
& & + {\cal H}_3
    - \bar{\mu}_{\rm A} \sum_i c_i^{\rm A}
    - \bar{\mu}_{\rm B} \sum_i c_i^{\rm B}  \; .
\label{eq1}
\end{eqnarray}
Here $c_i^{\rm X}$$\in$\{0,1\}
is the local occupation variable for species X
(X=A or~B), and the third adsorption state
(``empty'' or ``solvated'')
corresponds to $c_i^{\rm A}$=$c_i^{\rm B}$=0.
The sums $\sum_{\langle ij \rangle}^{(n)}$ and $\sum_i$
run over all
$n$th-neighbor bonds and over all adsorption sites,
respectively, $\Phi_{\rm XY}^{(n)}$ denotes the
effective XY pair interaction through an
$n$th-neighbor bond, and $\sum_n$ runs over the interaction ranges.
The term ${\cal H}_3$ contains three-particle \cite{EINS91} and
possibly multi-particle interactions. Both the interaction ranges and
the absence or presence of multi-particle interactions depend on the
specific system.
The change in electrochemical potential when one X
particle is removed from the bulk solution and
adsorbed on the surface
is $-\bar{\mu}_{\rm X}$.
The sign convention is such that $\Phi_{\rm XY}^{(n)}$$>$0
denotes an effective attraction, and $\bar{\mu}_{\rm X}$$>$0
denotes a tendency for adsorption in the absence of lateral
interactions.

The main differences
between models for particular systems are the binding-site
geometries of the
adsorbed species and the strengths of the effective,
lateral interactions.
(Straightforward modifications of Eq.~(\ref{eq1})
are necessary if the adsorption sites for the two species
are different,
as they are, {\it e.g.}, in the model
describing urea on Pt(100).) Some previously studied models
that can be
defined by Eq.~(\ref{eq1}) or similar lattice-gas
Hamiltonians, are the one for urea on Pt(100)
\cite{RIKV92,RIKV93A,RIKV93B,GAMB93B,HIGH93,RIKV95,GAMB94},
the model developed by Huckaby and Blum for UPD of
copper on Au(111) in the presence of sulfate
\cite{HUCK90,BLUM91,HUCK91,HUCK91B,JZHA95A,JZHA95B,BLUM93,%
BLUM94A,BLUM94B,LEGA95},
and the standard three-state models with single-site
bonding, used in previous studies of
poisoning and enhancement in multicomponent adsorption
\cite{RIKV88B,COLL89,RIKV91A,RIKV91B,RIKV92}.
As illustrations of the lattices and interactions that
can be used, we show in Fig.~\ref{figLATa} the model used for
urea adsorption on Pt(100)
\cite{RIKV92,RIKV93A,RIKV93B,GAMB93B,HIGH93,RIKV95,GAMB94}
and in Fig.~\ref{figLATb} one
used for copper UPD on Au(111) \cite{JZHA95A,JZHA95B}.

The thermodynamic density conjugate to
the electrochemical potential
$\bar{\mu}_{\rm X}$ in Eq.~(\ref{eq1})
is the surface coverage by species X,
\begin{equation}
\label{eq1b}
\Theta_{\rm X} = N^{-1}\sum_i c_i^{\rm X} \;,
\end{equation}
where $N$ is the total number of surface unit cells in the system.
To connect the electrochemical potentials to the
bulk concentrations [X] and the electrode potential $E$, one has
(in the weak-solution approximation):
\begin{equation}
\label{eq2}
\bar{\mu}_{\rm X} = {\mu}_{\rm X}^0
+ RT \ln {[\rm X] \over [\rm X]^0} - z_{\rm X}FE \;,
\end{equation}
where $R$ is the molar gas constant,
$T$ is the absolute temperature, $F$ is Faraday's constant,
and the effective electrovalence of X is
$z_{\rm X}$. The quantities superscripted with a 0 are reference
values which contain the local binding energies to the surface.
They are generally temperature dependent due, among other
effects, to rotational and vibrational modes.

In the absence of diffusion and double-layer effects and in the
limit that the potential sweep rate d$E$/d$t$$\rightarrow$0
\cite{BARD80}, the voltammetric
current $i$ per unit cell of the surface
is the time derivative of the charge transported across
the interface during the adsorption/desorption process.
With a sign convention such that oxidation/anodic
currents are considered positive, this charge is
\begin{equation}
\label{eq2c}
q = -e(z_{\rm A} \Theta_{\rm A} + z_{\rm B} \Theta_{\rm B}) \;,
\end{equation}
where $e$ is the elementary charge unit.
Using partial differentiation involving the relation
between the electrode potential and the
electrochemical potentials, Eq.~(\ref{eq2}), as well as the
Maxwell relation
${\partial \Theta_{\rm A}}/{\partial
\bar{\mu}_{\rm B}}$=${\partial
\Theta_{\rm B}}/{\partial \bar{\mu}_{\rm A}}$, one obtains $i$
in terms of the lattice-gas response functions
${\partial \Theta_{\rm X}}/{\partial \bar{\mu}_{\rm Y}}$:
\begin{equation}
i = e {\cal F}
\left\{ z_{\rm A}^2 \frac{\partial \Theta_{\rm A}}{\partial
\bar{\mu}_{\rm A}}
+ 2 z_{\rm A} z_{\rm B}
    \frac{\partial \Theta_{\rm B}}{\partial \bar{\mu}_{\rm A}}
           + z_{\rm B}^2
    \frac{\partial \Theta_{\rm B}}{\partial \bar{\mu}_{\rm B}}
    \right\} \frac{{\rm d}E}{{\rm d}t} \;.
\label{eq2b}
\end{equation}

It must be emphasized
that the interactions in Eq.~(\ref{eq1}) are {\it effective\/}
interactions mediated through several channels.
The mechanisms involved
include interactions between the adsorbate and the
substrate electron structure
\cite{EINS73,LAU78,EINS78,MUSC86,FEIB89},
adsorbate-induced local deformations of the substrate,
interactions with the fluid electrolyte
\cite{BLUM90,HUCK90,BLUM91,HUCK91,HUCK91B,RIKV91A,BLUM93,%
BLUM94A,BLUM94B,LEGA95}, and (screened) electrostatic interactions
\cite{GLOS93A}.
All these effects give rise to indirect,
effective interactions between
the adsorbate particles. In general one must assume that these
quantities could be dependent on temperature and
electrode potential.
The spatial structure of the generalized pair
interactions generally
involves rather complicated dependences on both
the magnitude and the direction of the vector joining the two
adsorbate particles, as well as on the
relative orientation of the particles.
Empirical models for the electronic contribution to the effective,
lateral pair interactions are well known
\cite{EINS73,LAU78,EINS78,MUSC86,FEIB89} and
are often of a decaying, oscillatory form proportional to
$\cos(2k_{\rm F}r)/r^\alpha$, where $k_{\rm F}$
is the Fermi momentum and
$\alpha$ may be between 2 and 5, depending
on the substrate's electronic
structure \cite{LAU78,EINS78}. However,
changes in the effective
interaction energies of only a few percent
may cause very substantial
changes in the finite-temperature phase diagram (see, {\it e.g.},
Refs.~\cite{RIKV93D,CCAG90,HILT92}). First-principles
calculations of lateral adsorbate interactions  to this level of
accuracy are not yet feasible,
even for the electronically mediated contributions \cite{FEIB89}.

Here we advocate an approach to the problem of determining
the effective adsorbate--adsorbate interaction energies, which
provides a practical alternative to the ideal ``first-principles''
approach mentioned above. This strategy consists in
fitting the thermodynamic and structural predictions
of the lattice-gas model directly to experiments, taking
into account as
wide a spectrum of experimental information as possible.
Obviously, this method also involves
considerable difficulties. In particular, the number of parameters
that can reasonably be included in
a lattice-gas model is large, and there is no {\it a priori}
guarantee that a minimal set of fitted interactions is unique.
Nevertheless, the encouraging results of
previous lattice-gas studies
of electrochemical systems that have employed this strategy
\cite{RIKV88A,RIKV88B,COLL89,BLUM90,HUCK90,BLUM91,HUCK91,%
HUCK91B,ARMA91A,RIKV91A,RIKV91B,RIKV92,RIKV93A,RIKV93B,%
GAMB93B,HIGH93,RIKV95,JZHA95A,JZHA95B,BLUM93,BLUM94A,%
BLUM94B,LEGA95}
indicate that when proper attention is paid to including
all available experimental information in a consistent fashion, the
predictive power of this approach is considerable. Furthermore,
as effective interactions obtained by first-principles calculations
become available in the future, the results obtained from
lattice-gas models
will provide crucial information for testing the consistency of
such first-principles interactions with the experimentally observed
thermodynamic and structural information.
The steps in the modeling strategy outlined here
can be summarized as follows.\\
1. Use prior theoretical and experimental knowledge
about the adsorbate
lattice structure and lattice constant and the shapes and
sizes of the
adsorbate particles to formulate a specific lattice-gas model.
Examples are shown in Figs.~\ref{figLATa} and~\ref{figLATb}.\\
2. Use available experimental information about
adsorbate coverages
and adlayer structure to determine the adsorbate phases
or at least
narrow down the possible choices as much as possible.\\
3. Perform a group-theoretical ground-state calculation
\cite{DOMA78,DOMA79,SCHI81} to determine a
minimal set of effective interactions compatible
with the observed
adsorbate phases. Relations between the
effective interactions
take the form of a set of inequalities
\cite{RIKV88A,RIKV88B,COLL89,RIKV91A,RIKV91B}.
A ground-state diagram (zero-temperature
phase diagram) is obtained by pairwise equating
the ground-state energies of the different phases. Examples of
ground-state diagrams corresponding to the specific models in
Figs.~\ref{figLATa} and~\ref{figLATb} are shown in
Figs.~\ref{figGSa} and~\ref{figGSb}, respectively.\\
4. At nonzero temperatures,
the thermodynamic and structural properties of
the lattice-gas model
constructed through steps 1--3 can be studied
by a number of
analytical and numerical methods, depending
on the quantities of
interest and the complexity of the Hamiltonian.
These methods include mean-field
approximations \cite{ARMA91A,HOMM89}
(although these can be unreliable for low-dimensional
systems with
short-range interactions \cite{RIKV93D}),
Pad{\'e}-approximant methods based on liquid theory
\cite{HUCK90,BLUM91,HUCK91,HUCK91B,BLUM93,BLUM94A,%
BLUM94B,LEGA95},
numerical TM calculations
\cite{RIKV88A,RIKV88B,COLL89,RIKV91A,RIKV91B,RIKV92},
and MC simulations
\cite{COLL89,RIKV93A,RIKV93B,GAMB93B,HIGH93,RIKV95%
,JZHA95A,JZHA95B}.\\
5. Whatever method is used to calculate the finite-temperature
properties
of the model, these should be used to refine the effective
interactions by comparison with the available experiments,
or by obtaining additional experimental data for
such comparison.\\
Steps 4 and 5 should be iterated
until satisfactory agreement between model and
experiment is achieved.

One of the main reasons for the rapid expansion
in theoretical surface
science over the last three decades is the development of numerical
methods that allow nonperturbative calculations of
thermal and structural
properties of statistical-mechanical systems. Two such methods,
which are
particularly well suited to the study of lattice-gas models,
are Monte Carlo (MC) simulation
\cite{COLL89,RIKV93A,RIKV93B,GAMB93B,HIGH93,RIKV95,%
JZHA95A,JZHA95B,BIND86,BIND92,BIND92B}
and numerical transfer-matrix (TM) calculations
\cite{RIKV88A,RIKV88B,COLL89,RIKV91A,DOMB60,HUAN63,STAN71,NIGH90}.
In combination with finite-size scaling analysis of
phase-transition phenomena \cite{BIND92,NIGH90,PRIV90},
these methods have
contributed significantly to the  theoretical understanding
of fluctuations and ordering at surfaces and interfaces.
The reason for our emphasis on non-perturbative numerical
methods is that they are much more accurate for
two-dimensional systems than even
quite sophisticated mean-field approximations \cite{RIKV93D},
yet they are quite easy to program.
Moreover, with modern computer technology their
implementation is well within the
resources of most researchers.

At present, a large number of monographs and textbooks
exist that describe
MC methods in great detail \cite{BIND86,BIND92,BIND92B}.
We therefore limit
ourselves to pointing out that these methods can
produce thermodynamic
and structural information for a variety of systems,
with a very
modest amount of programming and with computational
resource needs that
are readily met by modern workstations.
For example, all the
MC results presented here were obtained on workstations. For
studies of real
systems, MC models have the advantage that programs are
relatively easy
to modify to accommodate changes in lattice structure
and/or interaction geometries and ranges.

Despite their power and beauty, TM methods are much
less known outside
the statistical-mechanics community. However, good reviews are
available \cite{DOMB60,NIGH90}, and simple textbook expositions for
the one-dimensional case are quite illustrative
\cite{HUAN63,STAN71}. An abundance of details are scattered
throughout the technical literature and can be found,
together with further references in {\it e.g.}\
Refs.~\cite{RIKV88A,RIKV88B,COLL89,RIKV91A,RIKV84,AUKR90}.
Briefly, the method allows the
numerical calculation of free energies (an advantage
over MC, which does not
easily produce entropies), thermodynamic densities,
and their associated
response functions from the eigenvalues and eigenvectors of a
matrix of Boltzmann factors, called the transfer matrix.
In addition to the
ability to easily calculate free energies,
the method has the further
advantage over MC that the results are obtained
without statistical
errors. The main disadvantages, relative to MC,
are the limited system
sizes and interaction ranges that can be attained.
The first problem can
relatively easily be overcome with finite-size scaling.
The second, however,
severely restricts the applicability of TM methods to realistic
electrochemical systems.

\section{Poisoning and Enhancement Effects}
\label{secPE}

Depending on the relative interaction ranges and
strengths, the lattice-gas models discussed in Sec.~\ref{secMOD}
allow many topologically different adsorbate
phase diagrams. The specific
coadsorption phenomena, such as poisoning
\cite{RIKV88A,RIKV88B,COLL89,ARMA91A,RIKV91B,RIKV92,HOMM89}
or enhanced adsorption
\cite{RIKV88B,ARMA91A,RIKV91A,RIKV91B,RIKV92,HOMM89,KRUK95},
which occur for any particular set of interactions,
depend crucially on the detailed topology of the phase diagram
\cite{RIKV88A,RIKV88B,COLL89,RIKV91A,RIKV91B,RIKV92}.
The terms ``poisoning'' and
``enhancement'' can be defined as follows.\\
{\it Poisoning of {\rm A} by {\rm B}:}
When $\bar{\mu}_{\rm B}$ is increased at constant
$\bar{\mu}_{\rm A}$,
the total coverage, $\Theta_{\rm A} + \Theta_{\rm B}$,
goes through a minimum as
$\Theta_{\rm A}$ decreases sharply with only a small corresponding
increase in $\Theta_{\rm B}$.\\
{\it Enhancement of {\rm A} by {\rm B}:}
$\Theta_{\rm A}$ goes through a maximum as
$\bar{\mu}_{\rm B}$ is increased at constant
$\bar{\mu}_{\rm A}$. For
large $\bar{\mu}_{\rm B}$, the
enhancement gives way to substitutional desorption, each adsorbed
B particle replacing one or more A particles.

{For} both poisoning and enhancement, a measure of the
modification strength is the differential coadsorption ratio,
${\rm d}\Theta_{\rm A} / {\rm d}\Theta_{\rm B}$.
The modification is characterized as
{\it strong} if $|{\rm d}\Theta_{\rm A} /
{\rm d}\Theta_{\rm B}| > Z$,
the lattice coordination number.
In Refs.~\cite{COLL89,RIKV91A,RIKV91B} it was discussed in detail
how the modification strength is related to the interaction constants
for specific, triangular lattice-gas models through the shape of
the adsorbate phase diagram.
 From the standpoint of statistical mechanics,
strong modification
results from fluctuations typical of the region near
a line of critical end points, which joins a surface of
discontinuous phase
transitions to one of continuous transitions. More intuitively,
these fluctuations can be described as follows. For
poisoning, they correspond to an almost bare surface, from which the A
particles are repelled by a very small coverage of
repulsively interacting B
particles \cite{COLL89}. In the case of enhanced adsorption, the
corresponding picture is that of a surface almost fully covered by a
monolayer of A particles, which is ``pinned
down'' by a low concentration of attractively interacting
B particles \cite{RIKV91A}.
These considerations lead to inequalities that must
be obeyed by the interaction constants,
in order for the system to exhibit
either poisoning or enhancement of various strengths.
The inequalities are illustrated in
Fig.~1 of Ref.~\cite{RIKV91B} for triangular models with
nearest-neighbor interactions.

\subsection{An Example of Poisoning}
\label{secPOIS}

An example of poisoning is provided by the
model for the coadsorption of sulfur and hydrogen on Pt(111)
in acid aqueous environment, studied by Rikvold and coworkers
\cite{RIKV88A,RIKV88B,COLL89}.
In this case, the effective nearest-neighbor
lateral interactions
were obtained from experimental thermodynamic and
scattering data. The numerical
adsorption isotherms (obtained by both MC and TM methods)
gave maximum desorption ratios
d$\Theta_{\rm H}/$d$\Theta_{\rm S} \approx -7 \pm 1$,
in favorable agreement with experiments \cite{PROT86}.
It was argued that the general shape of the phase
diagram for this model
is characteristic of strong poisoning behavior \cite{COLL89}.

\subsection{An Example of Enhancement}
\label{sec_EA}

An application of lattice-gas models to study enhanced adsorption was
given by Rikvold and Deakin \cite{RIKV91A}, who analyzed
experimental data for the electrosorption of
organics on metal electrodes:
naphthalene on copper \cite{BOCK64B}
and n-decylamine on nickel \cite{BOCK64A}. They followed
a suggestion by Damaskin {\it et al.\/} \cite{DAMA71} that the
potential dependence of adsorption of organics on metals
can be attributed to the influence of coadsorbed hydrogen. Although
the experimental results
concerned rough, polycrystalline electrodes, a
simple nearest-neighbor model on a triangular lattice was used,
aiming merely for semiquantitative agreement.
The effective electrovalences
were taken as $z_{\rm H}$=+1 and $z_{\rm organic}$=0, and
the three effective interaction constants,
$\Phi_{\rm XY}$, together with
$\mu_{\rm H}^0$ and $\mu_{\rm organic}^0$, were determined by
nonlinear least-squares fits of numerical coadsorption
isotherms obtained from a TM calculation to the experimental data.
The experimental and fitted numerical
adsorption isotherms for naphthalene on copper are
shown in Fig.~\ref{figNAPH}.
The maxima are due to the formation of a
mixed naphthalene/hydrogen adsorbed phase in the potential region
between $-$1000 and $-$800 mV versus
the normal hydrogen electrode (NHE). The fitted lattice-gas
interactions are consistent with independent estimates
\cite{BOCK64B,BOCK64A},
as discussed in detail in Ref.~\cite{RIKV91A}.

\section{Adsorption on Single-Crystal Surfaces}
\label{secEX}

A major source of uncertainty in the applications
of simple lattice-gas
models to the experimental results discussed in the
previous section,
is the poor characterization of the electrode surfaces.
To remedy this
situation, Wieckowski and Rikvold with collaborators
have undertaken a series of studies of the electrosorption
of small molecules and ions on well-characterized
single-crystal surfaces.
A characteristic aspect of these systems is the high
specificity of the
adsorption phenomena with respect to the structures
of the substrate
lattice and the main adsorbate. A good geometric fit promotes
the formation of ordered adsorbate phases commensurate
with the substrate, which can be observed both by {\it in situ}\
atomic-scale spectroscopies and by {\it ex situ}\
scattering techniques.
The detailed experimental results that can be extracted from
such systems merit the construction of more complicated models with
longer-ranged and multi-particle interactions.

By way of examples we discuss two specific single-crystal
adsorption systems:
the electrosorption of urea on Pt(100) from an acid electrolyte
\cite{RIKV92,RIKV93A,RIKV93B,GAMB93B,HIGH93,RIKV95,GAMB94}
and the UPD
of copper on Au(111) from a sulfate-containing electrolyte
\cite{HUCK90,BLUM91,HUCK91,HUCK91B,JZHA95A,JZHA95B,BLUM93,%
BLUM94A,BLUM94B,LEGA95}.
Both systems exhibit a dramatic peak sharpening
in the cyclic voltammogram (CV),
from several hundred~mV to on the order of 10$\,$mV
when a small concentration of
the adsorbate species (urea,
or a mixture of sulfate and copper ions, respectively) is added to the
supporting electrolyte.
This effect is also exhibited by other systems, such as sulfuric
acid on Rh(111) \cite{RIKV95,GAMB94,JZHA95C}.
Whereas the urea/Pt(100) system develops only a single,
sharp CV peak \cite{RUBE91}, in the case of copper UPD,
two peaks, approximately 100~mV apart,
are exhibited \cite{KOLB87,ZEI87}.
We associate these effects with phase transitions in
the layer of contact adsorbed particles. These transitions involve the
replacement of a monolayer of adsorbed
hydrogen or copper on the negative-potential side of the CV peaks by
ordered submonolayers at more positive potentials.
The observed voltammetric changes
are much weaker or absent when the same
substances are adsorbed onto other crystal planes of the same
metals \cite{GAMB94,RUBE91,SCHU76}.
The high specificity with respect to the
adsorbent surface structure indicates that the effects
depend crucially on the geometric fit between (at least one of)
the adsorbate species and the surface. This observation was used in
developing the specific lattice-gas models.

\subsection{Urea on Pt(100)}
\label{secUPT}

In addition to the surface-specific
narrowing of the CV peak upon the addition
of urea to the supporting electrolyte, the experimental
observations to which the model was fitted are as follows.
(For details, see Refs.~\cite{GAMB93B,RIKV95}.)\\
1. The urea coverage $\Theta_{\rm U}$, measured {\it in situ\/}
by a radiochemical method
(RCM), changes over a potential range of approximately 20$\,$mV around
the CV peak position from near zero on the negative side to
approximately 1/4 monolayers (ML) on the positive side.\\
2. {\it Ex situ\/} Auger electron spectroscopy (AES)
studies are consistent with the RCM results.\\
3. {\it Ex situ\/} LEED studies at potentials on the
positive side of the CV peak
show an ordered c(2$\times$4) adsorbate structure,
consistent with an
ideal coverage of 1/4$\,$ML. Upon emersion on the negative side of
the CV peak,
only an unreconstructed (1$\times$1) surface is found.\\

The lattice-gas model developed to account for these observations
was based on the assumption that urea [CO(NH$_2)_2$]
coordinates the platinum through its nitrogen atoms
(or NH$_2$ groups), with the C=O group pointing away from the surface.
Since the unstrained N-N distance in urea matches the lattice constant
of the square Pt(100) surface quite well
(2.33~{\AA} \cite{ITAI77} versus 2.77~{\AA} \cite{KITT86}),
it was assumed that urea occupies two adsorption
sites on the square Pt(100) lattice.
Integration of the CV profiles indicates that the hydrogen saturation
coverage
in the negative-potential region corresponds to one elementary charge
per Pt(100) unit cell, and that most of the surface hydrogen is
desorbed in the same potential range where urea becomes adsorbed.
Therefore, it was assumed \cite{RIKV92} that hydrogen adsorbs in
the same on-top positions as the urea nitrogen atoms. This assumption
was recently strengthened by visible-infrared sum
generation spectroscopy
observations \cite{TADJ94,TADJ95}.
The resulting model \cite{RIKV92} is a dimer-monomer model
in which hydrogen is adsorbed at the nodes and urea on
the bonds of a square
lattice representing the Pt(100) surface. Simultaneous occupation
of bonds that share a node
by two or more urea molecules
is excluded, as is occupation by
hydrogen of a node adjacent to a bond occupied by urea.
In order to stabilize the observed c(2$\times$4) phase, effective
interactions were included through eighth-nearest neighbors
\cite{RIKV93A,RIKV93B,GAMB93B,HIGH93,RIKV95,GAMB94}.
The configuration energies are given by
Eq.~(\ref{eq1}) with A=U (urea) and B=H (hydrogen).
The model is illustrated in Fig.~\ref{figLATa} and its
ground-state diagram in Fig.~\protect\ref{figGSa}.
The effective lattice-gas interactions
were determined from ground-state
calculations followed by numerical MC simulations.

The numerical simulations, which used systems with up to
32$\times$32 square-lattice unit cells,
were performed with a heat-bath MC algorithm
\cite{BIND86,BIND92,BIND92B}
with updates of clusters
consisting of five nearest-neighbor nodes arranged in a
cross, plus their four connecting bonds. After symmetry
reductions these
clusters have 64 different configurations, and the
corresponding code is
rather slow in terms of machine time per MC step.
However, the additional transitions allowed by these
clusters, relative
to minimal clusters consisting of two nodes and their
connecting bond,
include ``diffusion-like'' moves in which the urea molecules can go
from one bond to another and the hydrogen atoms from one node to
another, without changing the local coverages within the cluster.
These moves significantly
reduce the free-energy barriers that must be surmounted in
order to locally minimize the adsorbate free energy, and they
dramatically reduce the number of MC steps per site (MCSS) necessary
for the system to reach thermodynamic equilibrium. For this system,
simulated ``LEED patterns'' were obtained as the squared Fourier
transform of the adsorbed urea configurations. These were obtained by
the Fast Fourier Transform algorithm
and averaged in the same way as the thermodynamic quantities
\cite{RIKV95}.

Since the number
of model parameters is large, the numerical calculations
are time consuming, and the experimental data
concern a number of different quantities, parameter estimation
by a formal optimization procedure was not a practical alternative
for this study.
(This contrasts with the simpler situations discussed in
Refs.~\cite{RIKV91A,RIKV93D,HILT92},
where a small number of lattice-gas parameters
could be determined by a formal least-squares procedure to fit
extensive experimental results for a single thermodynamic quantity.)
To make maximum use of all available information,
the model parameters were therefore varied
``by hand'', taking into consideration both the various experimental
results and available chemical and physical background information,
until acceptable
agreement was obtained with room-temperature experimental results. In
particular, agreement was  sought between
the shapes of the simulated and experimental CV profiles,
as shown in in Fig.~\ref{figCVa}.
The resulting interactions are given in the
caption of Fig.~\ref{figLATa}.

\subsection{Copper UPD on Au(111)}
\label{secUPD}

Underpotential deposition (UPD) is a process whereby a monolayer
or less of one metal is electrochemically adsorbed onto
another in a range of
electrode potentials more positive than those where bulk
deposition would
occur \cite{BARD80}.
The UPD of copper on Au(111) electrodes in
sulfate-containing electrolytes has been intensively studied,
both experimentally (see discussion of the
literature in Ref.~\cite{JZHA95B}) and theoretically
\cite{HUCK90,BLUM91,HUCK91,HUCK91B,JZHA95A,JZHA95B,BLUM93,%
BLUM94A,BLUM94B,LEGA95}.
The most striking feature observed in CV experiments with Au(111)
electrodes in sulfate-containing electrolyte
is the appearance of two peaks,
separated by about 100$\sim$150 mV, upon the
addition of Cu$^{2+}$ ions
\cite{KOLB87,ZEI87}.
Typical CV profiles are shown in Fig.~\ref{figCVb},
together with
preliminary simulation results \cite{JZHA95A,JZHA95B}.
In the potential range between the peaks, the adsorbate
layer is believed to
have a ($\sqrt3$$\times$$\sqrt3$) structure consisting
of 2/3$\,$ML copper and 1/3$\,$ML sulfate
\cite{HUCK90,BLUM91,HUCK91,HUCK91B,JZHA95B,BLUM93,BLUM94A,%
BLUM94B,LEGA95,ZSHI94B,ZSHI94C,ZSHI95}.

The lattice-gas model for UPD of copper on Au(111)
in sulfate-containing
electrolyte, used in Refs.~\cite{JZHA95A,JZHA95B},
is a refinement of the model introduced and studied by
Huckaby and Blum
\cite{HUCK90,BLUM91,HUCK91,HUCK91B,BLUM93,BLUM94A,%
BLUM94B,LEGA95}.
It is based on the assumption that
the sulfate coordinates the triangular
Au(111) surface through three of its
oxygen atoms, with the fourth S-O bond pointing away
from the surface,
as is also the most likely adsorption geometry on Rh(111)
\cite{RIKV95}. This adsorption
geometry gives the sulfate a ``footprint''
in the shape of an approximately
equilateral triangle with a O-O distance of 2.4 \AA \,
\cite{PASC65}, reasonably matching
the lattice constant for the triangular Au(111)
unit cell, 2.88 \AA \,
\cite{KITT86}. The copper is assumed to compete
for the same adsorption sites as the sulfate.
The configuration energies are given by
Eq.~(\ref{eq1}) with A=S (sulfate) and B=C (copper).
The model is illustrated in Fig.~\ref{figLATb}
and its ground-state diagram in Fig.~\ref{figGSb}.

It has been experimentally observed \cite{ZSHI94C,ZSHI95}
that sulfate remains adsorbed on top of the copper monolayer
in the negative-potential region.
In principle, this system should therefore be
described by a multilayer lattice-gas model \cite{KRUK95}.
In Refs.~\cite{JZHA95A,JZHA95B} this complication was
avoided by using
the following, simple mean-field estimate for
the sulfate coverage in this second layer:
\begin{equation}
\label{eqTheta2}
\Theta_{\rm S}^{(2)} = \alpha\Theta_{\rm C}(1/3-\Theta_{\rm S}) \;,
\end{equation}
which allows the difference between the first-layer coverage
$\Theta_{\rm S}$
and its saturation value of 1/3 to be transferred to the top
of the copper
layer. The factor $\alpha$ is a phenomenological constant.
Since the transfer of sulfate between the gold and copper
surfaces does not
involve an oxidation/reduction process, the total
charge transport
per unit cell during the adsorption/desorption
process becomes
\begin{equation}
\label{eqQ2}
q = -e [ z_{\rm s} ( \Theta_{\rm S} + \Theta_{\rm S}^{(2)} )
+ z_{\rm C} \Theta_{\rm C} ] \;,
\end{equation}
giving a CV current density which reduces to that of
Eq.~(\ref{eq2b}) for $\alpha$=0:
\begin{eqnarray}
i & = & eF\left\{ z_{\rm S}^2(1-\alpha\Theta_{\rm C})
\left.\frac{\partial\Theta_{\rm S}}
{\partial\bar{\mu}_{\rm S}}\right|_{\bar{\mu}_{\rm C}}
+z_{\rm C}(z_{\rm C}-2\alpha z_{\rm S}\Theta_{\rm S}/3)
\left.\frac{\partial \Theta_{\rm C}}
{\partial \bar{\mu}_{\rm C}} \right|_{\bar{\mu}_{\rm S}}
\right.  \nonumber \\
  &   & +
\left.
z_{\rm S}(2z_{\rm C} +
\alpha z_{\rm S}(1/3-\Theta_{\rm S})-\alpha z_{\rm C}
\Theta_{\rm C})
\left.\frac{\partial\Theta_{\rm S}}
{\partial \bar{\mu}_{\rm C}}\right|_{\bar{\mu}_{\rm S}}
\right\}\frac{{\rm d}E}{{\rm d}t} \;.
\label{eqIalpha}
\end{eqnarray}
The effective electrovalences, $z_{\rm S}$ and $z_{\rm C}$,
must be determined from experiments
\cite{ZSHI94B,ZSHI94C,ZSHI95}.
In Refs.~\cite{JZHA95A,JZHA95B} the approximate values,
$z_{\rm C}$=+2 and $z_{\rm S}$=$-$2, were used.

The ground-state diagram corresponding to the interactions used
in Refs.~\cite{JZHA95A,JZHA95B} is shown in Fig.~\ref{figGSb}.
For large negative $\bar{\mu}_{\rm S}$,
only copper adsorption is possible,
and the phase diagram is that of the lattice-gas model
corresponding to the
triangular-lattice antiferromagnet with next-nearest
neighbor ferromagnetic
interactions \cite{LAND83}. Similarly, in the limit
of large positive
$\bar{\mu}_{\rm S}$ and large negative
$\bar{\mu}_{\rm C}$, the
zero-temperature phase is the
$(\sqrt{3}\!\times\!\sqrt{3})_0^{1/3}$ sulfate
phase characteristic of the hard-hexagon model
\cite{BLUM91,HUCK91,HUCK91B,BLUM93,%
BLUM94A,BLUM94B,LEGA95,BAXT82}.
The phase diagram for intermediate electrochemical
potentials is quite complicated.

To obtain adsorption isotherms and CV currents at room temperature,
MC simulations were performed on a 30$\times$30 triangular lattice,
using a heat-bath algorithm \cite{BIND86,BIND92,BIND92B}
with updates at randomly chosen sites. In order to avoid
getting stuck in
metastable configurations (a problem which is exacerbated by the
nearest-neighbor sulfate-sulfate exclusion), clusters
consisting of two nearest-neighbor sites were updated simultaneously.

The potential scan path corresponding to the CV shown in
Fig.~\ref{figCVb} is indicated by the dotted line labeled ``1''
in the ground-state diagram, Fig.~\ref{figGSb}.
With the aid of this diagram, it is easy to analyse the
simulation results.
As was pointed out above, there is experimental evidence
that at the negative end of the UPD potential range, sulfate
adsorbs in a neutral submonolayer on top of the
monolayer of copper, with a
coverage $\Theta_{\rm S}^{(2)}\!\approx\! 0.2$ \cite{ZSHI94C,ZSHI95}.
This corresponds to $\alpha \! = \! 0.6$ in Eq.~(\ref{eqTheta2}),
which was used to obtain the simulated CV current
shown in Fig.~\ref{figCVb}.
Starting from the negative end, we scan in the
direction of positive electrode potential (upper left
to lower right in
Fig.~\ref{figGSb}). Near the CV peak at approximately 70 mV, the
sulfate begins to compete with copper for the
gold surface sites, resulting
in a third of the copper desorbing into the bulk and
being replaced by
sulfate. The potential range over which the replacement takes
place corresponds to a peak width of about 30 mV. Due to the
strong effective
attraction between the copper and sulfate adparticles, a mixed
$(\sqrt{3}\!\times\!\sqrt{3})_{2/3}^{1/3}$ phase is formed, which
extends through the entire potential region between the two CV peaks.
As the CV peak at approximately 170 mV is reached,
most of the copper is desorbed within
a potential range of about 20 mV. As it is thus deprived of the
stabilizing influence
of the coadsorbed copper, the sulfate is partly
desorbed, reducing
$\Theta_{\rm S}$ from 1/3 to approximately 0.16.
This system provides another illustrative example of the
enhanced adsorption phenomenon described in Sec.~\ref{sec_EA}.
The $(\sqrt{3}\!\times\!\sqrt{7})_0^{1/5}$ phase found in the
potential region
near 200 mV is consistent with experimental observations
on copper free systems \cite{MAGN90,EDEN94}.
Eventually, more positive electrode potentials
cause the sulfate to form its saturated
$(\sqrt{3}\!\times\!\sqrt{3})_0^{1/3}$ hard-hexagon phase.
However, in the model, this transition occurs at a somewhat more
negative potential than is observed experimentally
\cite{ZSHI94B,ZSHI94C,ZSHI95}.
The scenario described here corresponds closely to that
proposed by Huckaby and Blum
\cite{HUCK90,BLUM91,HUCK91,HUCK91B,BLUM93,%
BLUM94A,BLUM94B,LEGA95}.
The agreement between the experimental and theoretical
results is reasonable,
except for large positive $E$, where the model predicts less
copper and more
sulfate on the surface than indicated by the experiments.
The heights of the CV peaks predicted by the model are larger
than what is observed in experiments, a discrepancy which
is probably due to defects on the
electrodes used in the experiments.

\section{Conclusion}
\label{secDIS}

We have briefly reviewed the application of
statistical-mechanical lattice-gas modeling to
specific adsorption in the double-layer region.
The method is well suited to describe ordering and fluctuation
effects in the contact-adsorbed layer, which are strongly
influenced by
effective, lateral interactions. Phenomena that can be described
include poisoning and enhancement effects, and concrete examples
were given for several systems of experimental interest.
The effective interactions arise from a
number of different sources, including mediation through the
substrate electrons, through phonons, and through the
fluid near the surface, and their calculation from first principles
is not yet feasible in general.
The alternative route advocated here provides a microscopic
picture of the adsorbate structure, as well as a
procedure for estimating
approximate effective interaction energies from
experimentally observed structural and
thermodynamic quantities.
The resulting models have considerable
predictive power regarding the dependences of observed thermodynamic
quantities on the electrochemical potential and the
bulk solute concentrations, as well as on the geometric
structure of the
substrate and the adsorbates.
Since the methods discussed are simple to program and
not particularly computationally intensive,
they are well suited for
experimental data analysis.
\vspace{0.25in}

\noindent {\bf Acknowledgement} \\*[0.15in]
We acknowledge useful discussions with L.~Blum.
Helpful comments on the manuscript were provided by
M.~A.\ Novotny,
R.~A.\ Ramos,
H.~L.\ Richards,
and
S.~W.\ Sides.
This research was supported by
the Florida State University (FSU) Supercomputer
Computations Research Institute
under US Department of Energy Contract No.\ DE-FC05-85ER25000,
by the FSU Center for
Materials Science and Technology, and
by the University of Illinois Frederick
Seitz Materials Research Laboratory
under US Department of Energy Contract No.\ DE-AC02-76ER01198.
Work at FSU was also supported by US National Science Foundation
Grant No.\ DMR-9315969.

%\bibliography{/a/alpha2/home/scri42/users/rikvold%
%/decstation/tex/biblio/elchem}
%\bibliographystyle{prsty}

\clearpage

\noindent{\bf Figure Captions}

\begin{figure}[ht]
\caption[]{
The lattice-gas model used
in Refs.~\protect\cite{RIKV93A,RIKV93B,GAMB93B,HIGH93,RIKV95}
to describe the coadsorption of urea (U) and hydrogen (H) on Pt(100).
The relative positions of hydrogen ($\bullet$)
and urea (filled rectangles) correspond to the
effective interactions in Eq.~(\protect\ref{eq1}),
which are invariant under symmetry operations on the lattice.
The values used in Refs.~\protect\cite{GAMB93B,HIGH93,RIKV95}
were (in kJ/mol)
$\Phi_{\rm HH}^{(1)}$=$-$2.0,
$\Phi_{\rm HU}^{(1)}$=$-$8.0,
$\Phi_{\rm HU}^{(2)}$=$-$4.0,
$\Phi_{\rm UU}^{(1)}$=$-$13.0,
$\Phi_{\rm UU}^{(2)}$=$-$10.0,
$\Phi_{\rm UU}^{(3)}$=$-$5.9,
$\Phi_{\rm UU}^{(4)}$=$-$0.5,
$\Phi_{\rm UU}^{(5)}$=$-$2.5,
$\Phi_{\rm UU}^{(6)}$=$-$3.0,
$\Phi_{\rm UU}^{(7)}$=$+$0.25,
$\Phi_{\rm UU}^{(8)}$=$-$2.0, and the effective electrovalences were
taken as $z_{\rm H}$=+1 and $z_{\rm U}$=$-$1.
After Ref.~\protect\cite{RIKV95}.
}
\label{figLATa}
\end{figure}

\begin{figure}[ht]
\caption[]{
The lattice-gas model used in Refs.~\protect\cite{JZHA95A,JZHA95B}
to describe the UPD of copper (C) on Au(111)
in the presence of sulfate (S). The relative positions of copper
($\bullet$) and sulfate ($\bigtriangleup$)
correspond to the effective interactions in Eq.~(\protect\ref{eq1}),
which are invariant under symmetry operations on the lattice.
The numbers are the corresponding values
of $\Phi_{\rm XY}^{(l)}$ used in
Refs.~\protect\cite{JZHA95A,JZHA95B}, given in kJ/mol.
After Ref.~\protect\cite{JZHA95B}.
}
\label{figLATb}
\end{figure}

\begin{figure}[ht]
\caption[]{
Ground-state diagram for the lattice-gas model of
urea on Pt(100), shown in the
($\bar{\mu}_{\rm U} , \bar{\mu}_{\rm H}$) plane.
The model parameters are the same as in Fig.~\protect\ref{figLATa}.
(a) The zero-temperature phase boundary between the c(2$\times$4)
phase with
$\Theta_{\rm U}$=1/4, indicated as c(2$\times$4)U in the figure,
and the (1$\times$1) phase with $\Theta_{\rm H}$=1, indicated as
(1$\times$1)H, is shown as a solid line
together with the electrochemical potentials corresponding to
room-temperature experimental
($\times$ connected by dotted lines) and simulated
(+ connected by dashed lines) CV peak positions.
The solid arrows represent positive-going $E$ scans from $-$106$\,$mV
to $-$56$\,$mV versus Ag/AgCl at room temperature.
 From left to right they represent [U]=2.0, 1.0, and 0.5$\,$mM.
Simulated and experimental CV currents along the scan at 1.0$\,$mM
are shown in Fig.~\protect\ref{figCVa}.
(b) A full ground-state diagram for the interactions used,
showing all the phases present. The phase indicated in the figure by
c(2$\times$4)UU has $\Theta_{\rm U}$=1/2, (1$\times$3)U has
$\Theta_{\rm U}$=1/3, ($\sqrt 2$$\times$$\sqrt 2$)H has
$\Theta_{\rm H}$=1/2, and (1$\times$1)0 is the empty lattice.
The phase regions outside the dotted box, which corresponds
to panel (a), are not believed to be experimentally relevant.
After Ref.~\protect\cite{RIKV95}.
}
\label{figGSa}
\end{figure}

\begin{figure}[ht]
\caption[]{
Ground-state diagram for the lattice-gas model of copper
UPD on Au(111)
in sulfuric-acid electrolyte, shown in the
($\bar{\mu}_{\rm S} , \bar{\mu}_{\rm C}$) plane.
The model parameters are the same as in Fig.~\protect\ref{figLATb}.
The solid lines represent zero-temperature phase boundaries,
and the dotted lines represent voltammetric scan paths at
room temperature. The scan path labeled ``1'' is fitted to an
experiment with an electrolyte containing
1.0 mM CuSO$_4$ \protect\cite{JZHA95B}, whereas
``2'' and ``3'' represent simulations corresponding to
5.0 mM and 0.2 mM
sulfate with 1.0 mM Cu$^{2+}$, respectively.
The end points of the dotted lines, marked $+$, correspond
to electrode potentials 55 mV (upper left) and 245 mV (lower right)
versus Cu/Cu$^+$, respectively.
The phases are indicated as
($X\!\times\!Y$)$_{\theta_{\rm C}}^{\theta_{\rm S}}$. The solid
squares indicate the left-hand peak positions,
and the solid diamonds indicate the right-hand peak positions
of simulated room-temperature CV currents, such as that shown in
Fig.~\protect\ref{figCVb}.
After Ref.~\protect\cite{JZHA95B}.
}
\label{figGSb}
\end{figure}

\begin{figure}[ht]
\caption[]{
Electrosorption of naphthalene on copper in alkaline
aqueous environment.
Experimental (data points connected by dotted
straight lines)~\protect\cite{BOCK64B}
and fitted numerical adsorption isotherms (solid curves)
\protect\cite{RIKV91A} are shown.
The lattice-gas parameters,
determined by a nonlinear least-squares
fit, are also given. From
below to above, the isotherms correspond to
naphthalene concentrations of 2.5, 5.0, 7.5, and
10.0$\times$10$^{-5}$~M.
After Refs.~\protect\cite{RIKV91A,RIKV91B,RIKV92}.
}
\label{figNAPH}
\end{figure}

\begin{figure}[ht]
\caption[]{
Cyclic voltammogram (CV) for urea adsorbed
on Pt(100) in 0.1$\,$M HClO$_4$ at room temperature.
Experimental (dashed curves) and simulated ($\Diamond$ and solid
curve) normalized CV currents,
$i$/(d$E$/d$t$) in elementary charges per mV per Pt(100) unit cell,
at 1.0$\,$mM bulk urea.
The two dashed curves are representative negative-going
voltammograms, and their differences indicate the experimental
uncertainty. The model parameters are
given in the caption of Fig.~\protect\ref{figLATa}.
After Ref.~\protect\cite{RIKV95}.
}
\label{figCVa}
\end{figure}

\begin{figure}[ht]
\caption[]{
Cyclic voltammogram (CV)
for copper UPD on Au(111) at room temperature,
corresponding to the scan path labeled ``1'' in
Fig.~\protect\ref{figGSb}.
Experimental (dashed curve) and preliminary simulated (solid curve),
normalized CV currents \protect\cite{JZHA95B}.
Left scale: CV current density. Right scale: normalized CV
current density,
$i$/(d$E$/d$t$), in electrons per mV per Au(111) unit cell.
The model parameters are given in Fig.~\protect\ref{figLATb}.
After Ref.~\protect\cite{JZHA95B}.
}
\label{figCVb}
\end{figure}

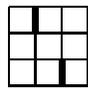
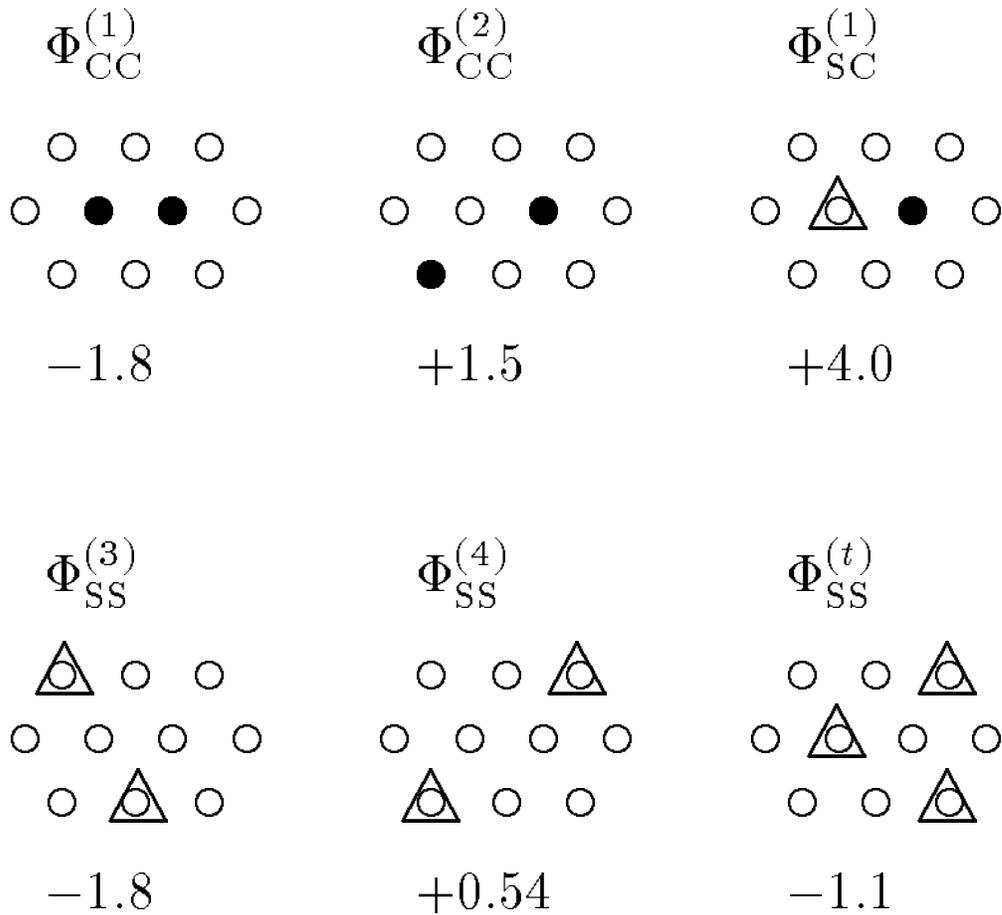
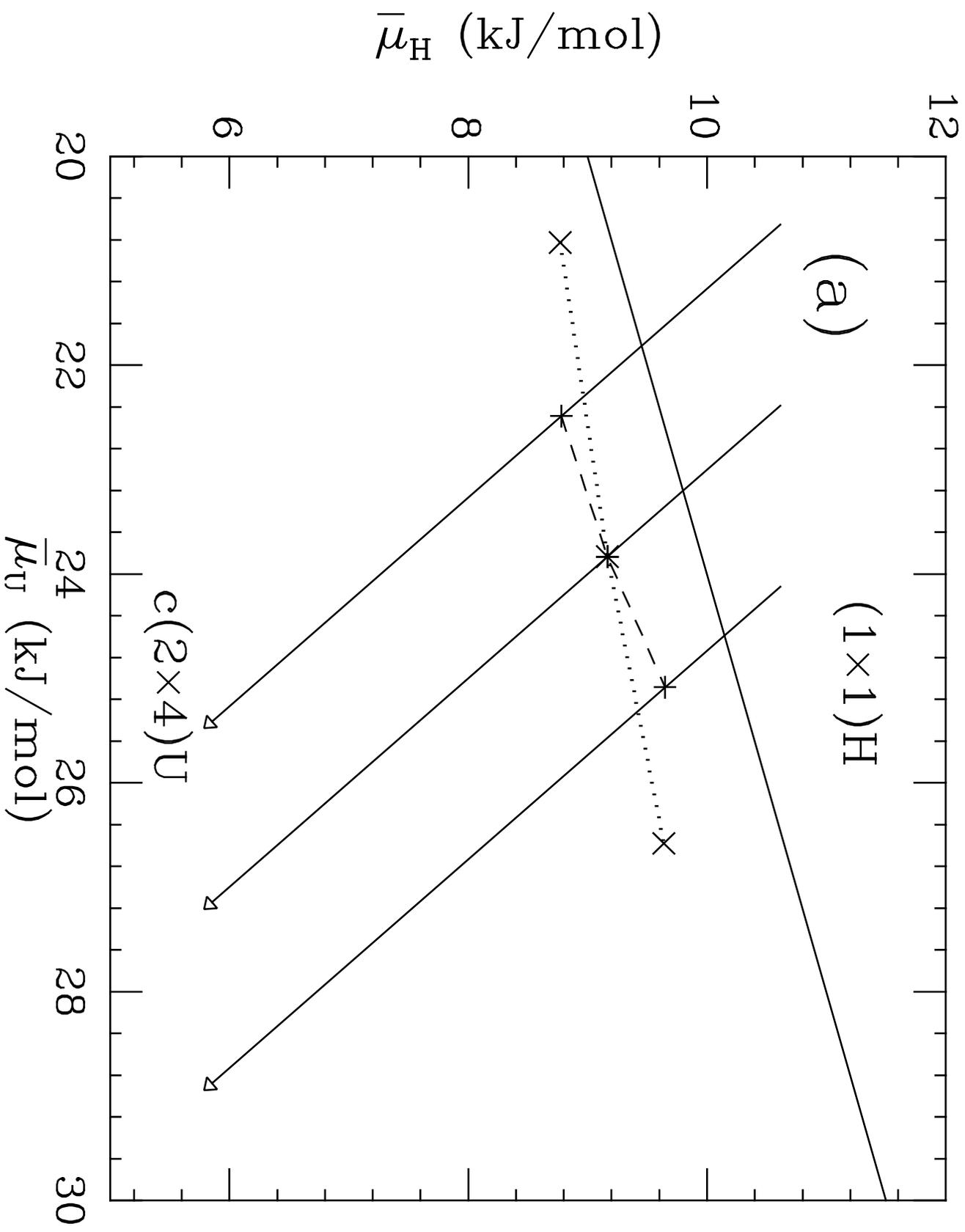
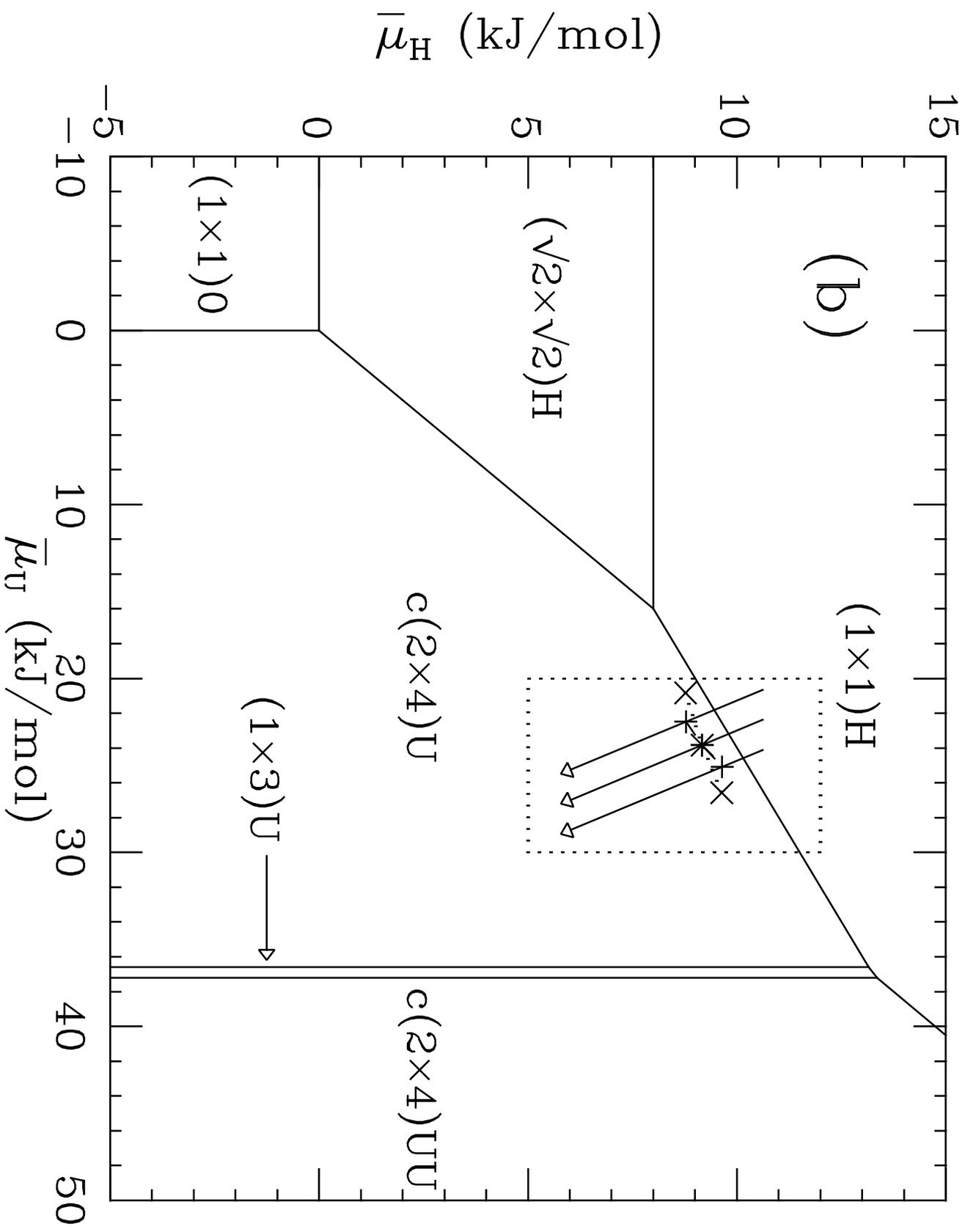
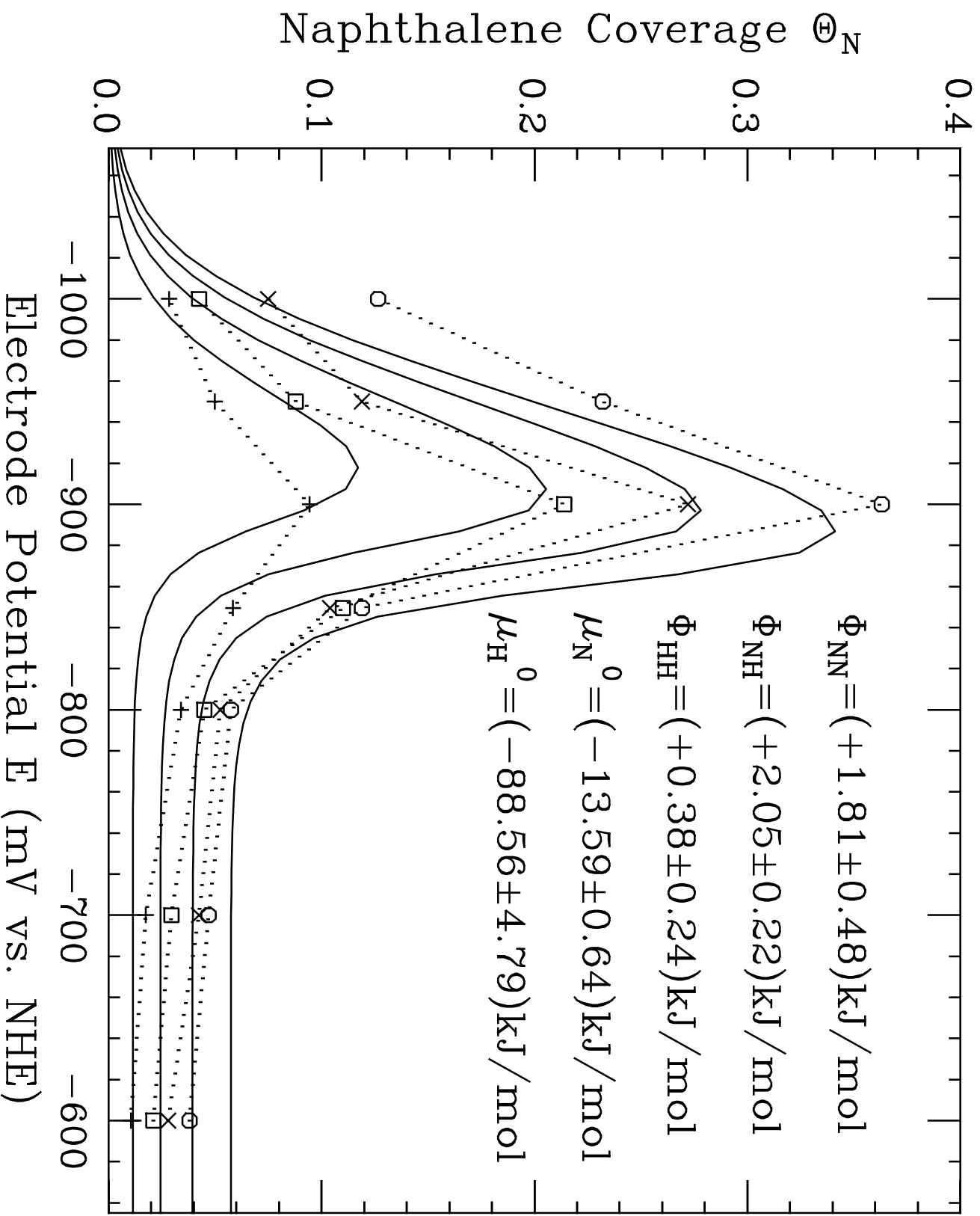
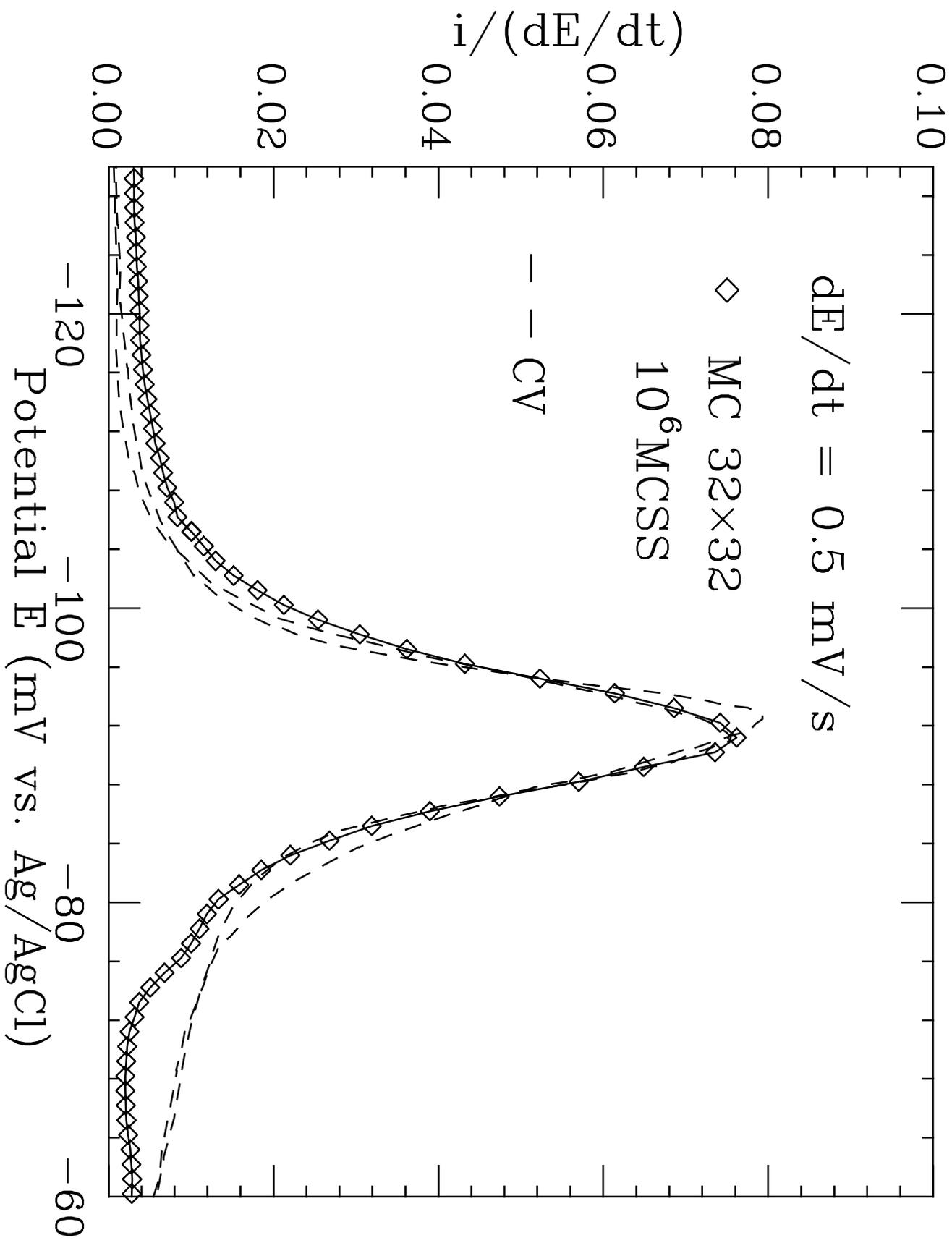
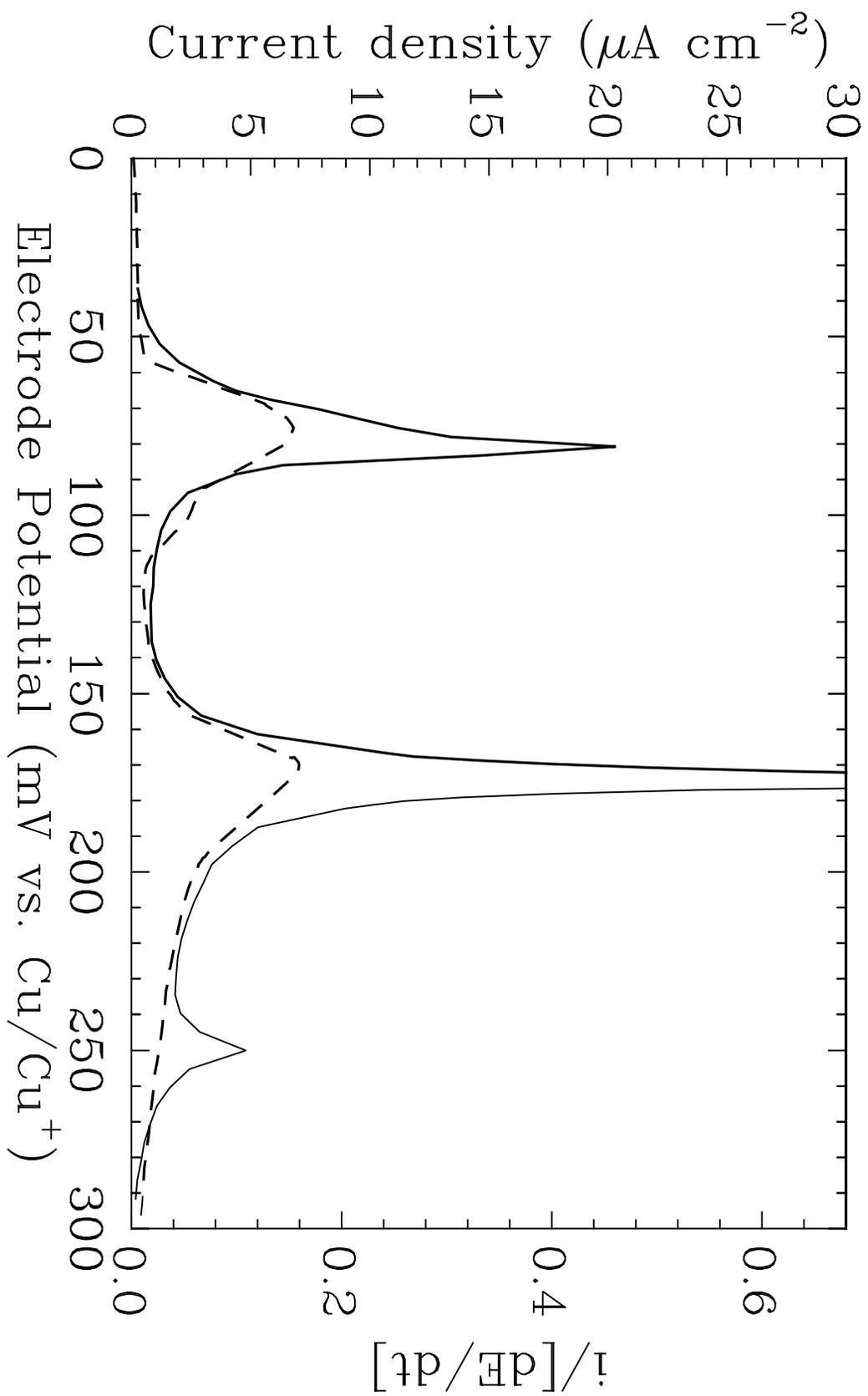
\end{document}